\documentclass[a4paper]{elsart}
\usepackage{graphicx}
\begin{document}

\journal{Journal of Computational Physics}

\begin{frontmatter}

\title{An efficient algorithm to generate large random uncorrelated Euclidean distances: the random link model}

\author{C\'esar Augusto Sangaletti Ter\c{c}ariol} 
\thanks{Centro Universit\'ario Bar\~ao de Mau\'a; Rua Ramos de Azevedo, 423 ; 
             14090-180, Ribeir\~ao Preto, SP, Brazil}
\and
\author{Alexandre Souto Martinez\corauthref{cor1}}
\corauth[cor1]{\ead{asmartinez@ffclrp.usp.br}}

\address{Faculdade de Filosofia, Ci\^encias e Letras de Ribeir\~ao Preto (FFCLRP) \\
         Universidade de S\~ao Paulo (USP) \\
         Av.~Bandeirantes, 3900 \\
         14040-901  Ribeir\~ao Preto, SP, Brazil. }

\date{\today}


\begin{abstract}
A disordered medium is often constructed by $N$ points independently and identically distributed in a $d$-dimensional hyperspace. 
Characteristics related to the statistics of this system is known as the random point problem.  
As $d \rightarrow \infty$, the distances between two points become independent random variables, leading to  its mean field description: the random link model. 
While the numerical treatment of large random point problems pose no major difficulty, the same is not true for large random link systems due to Euclidean restrictions. 
Exploring the deterministic nature of the congruential pseudo-random number generators, we present techniques which allow the consideration of models with memory consumption of order $O(N)$, instead of $O(N^2)$ in a naive implementation but with the same time dependence $O(N^2)$. 
\end{abstract}

\end{frontmatter}

\section{Introduction}

The random point problem (RPP) is a classical approach to construct disordered (random) media. 
In this problem, $N$ points are independently and identically distributed (i.i.d.) along the edges of a $d$-dimensional hypercube. 
Due to boundary effects and triangular restrictions, the distances between any two points are not all independent random variables. 
Periodic boundary conditions are frequently used to minimize the boundary effect.
As the system dimensionality increases, for fixed $N$ boundary effects become more and more pronounced and the distances become less and less correlated when periodic boundary conditions are used. 
As $d \rightarrow \infty$, all the two-point distances are i.i.d. random variables and this model is known as the \emph{random link (distance) model}~\cite{mezard:1986} (RLM).
In the RLM, there exist two Euclidean constraints: (i) the distance from a point to itself is always null ($D_{ii} = 0$, for all $i$) and (ii) the forward and backward distances are equal ($D_{ij} = D_{ji}$, for all $i$, $j$).
If both Euclidean constraints are relaxed, this model becomes the \emph{random map model}~\cite{harris:1960,derrida:2:1997}, which is the mean field approximation for Kauffman automata~\cite{kauffman:1969}.

Both, the RPP and the RLM (RPP mean field description) have been very fruitful in the determination of numerical and analytical results in several interesting systems.
These applications range from  statistics on the optimal trajectories in the context of traveling salesman problem on a random set of cities~\cite{percus:1996,percus:1997,percus:1999,aldous:2003,aldous:2005} passing by frustrated dimerization optimization modeled by the minimum matching problem~\cite{Boutet:1997,Boutet:1998} (or equivalently spin-glasses~\cite{Boutet:1997}) to partial self-avoiding deterministic~\cite{kinouchi:1:2002,tercariol_2005} and stochastic~\cite{risaugusman:1:2003,martinez:1:2004} walks. 
The high-dimensional case for these partial self-avoiding walks has been our main motivation to consider the RLM.  
In the deterministic walk~\cite{lima_prl2001,stanley_2001,boyer_2004,boyer_2005}, one is interested only on the neighborhood ranking of random points. Indeed, Euclidean distances are only a means to obtain this ranking, and this is independent of a particular distance probability distribution function (pdf) choices~\cite{tercariol_2005}. 
Here, we will consider only uniform distance pdf, nevertheless the algorithm can be easily adapted to treat any distance pdf, such as the one with the pseudo-dimension parameter~\cite{aldous:2003}.  

In the simplest congruential implementation, a numerical random-number generator is a deterministic algorithm initialized by a single integer variable $S_1$, called \emph{seed}.
At each generator call, this initial seed $S_1$ is deterministically modified and gives rise to the uncorrelated sequence of integers $S_2$,~$S_3$,~$\ldots$,  uniformly distributed in the interval $[0, 2^{m}-1[$, for any $m \ge 0$. 
After running all values in this interval, the seed reassume its initial value $S_1$ and the $m$-cycle reinitiates. 
For each integer seed value, the pseudo-random number generator commonly returns a  real number, which is uniformly distributed in the interval $[0,1[$, while keeping track of the seed value. 

In Sec.~\ref{sec_rpp}, the RPP is considered and its numerical implementation is discussed. 
Next, in Sec.~\ref{sec_rlm} we consider the RLM and we show that the straightforward implementation considering the distance matrix permits only small numerical systems simulations in a computer. 
Thus, we consider two alternative algorithms to implement the generation of random Euclidean distances by exploring the reproducibility of the pseudo-random number generators. 
Final remarks are addressed in Sec.~\ref{sec_conc}

\section{Random Point Problem}
\label{sec_rpp}

Consider a disordered medium made of $N$ points embedded in a $d$-dimensional Euclidean hyperspace.
The coordinates $x_i^{(k)}$ of these points are independent and randomly generated following a given common pdf $p_d(x)$ (for instance, uniform in a line segment of length $L$). 
The distance between any pair of points $i$ and $j$ is obtained by the Euclidean metrics:
\begin{equation}
D_{ij}^2 = \sum_{k=1}^d \left[ x_i^{(k)}-x_j^{(k)} \right] ^2 \; .
\label{eq:metrica_euclideana}
\end{equation}

A possible computational implementation of the RPP mainly consists of the following two steps: (i) randomly generate the coordinates $x_i^{(k)}$ and store them in a $N \times d$ matrix (\emph{coordinate matrix}) and (ii) use Eq.~\ref{eq:metrica_euclideana} to calculate the distance between any pair of points.
If one wishes to compare distances among points, one must store the distance values in a $N \times N$ matrix (\emph{distance matrix}) leading to $O(N^2)$ time consumption.
The declaration of the distance matrix corresponds to large computer memory consumption.
In numerical applications, this limits system sizes (typically to $N = 720$ in FORTRAN compilers and $N = 15000$ in C++).
Using an alternative procedure, one can avoid the distance matrix declaration.
This procedure consists to declare a vector of size $N$ (\emph{mask}) rather than the $N \times N$ distance matrix and calculate only the distances related to a given point at each time step (for instance, to determine its nearest neighbors).
Thus, the coordinate matrix, via the mask, saves us from the distance matrix declaration.
The only computational waste is to calculate the same distance ($D_{ij} = D_{ji}$) twice (it could have been calculated only once if one had the distance matrix).
Nevertheless, the time dependence is kept proportional to $N^2$. 

To minimize the boundary effect, it is important to consider periodic boundary conditions and to keep fixed the mean point separation ($\ell = \rho^{1/d} = L/N^{1/d}$, where $L$ is a typical system size and $\rho$ the point density), so that one has to increase $N$, as the system dimensionality increases.
Since high-dimensional systems are to be considered, even the declaration of the coordinate matrix may consume a lot of computer random access memory (RAM).
This introduces additional computational difficulties once the system size has found a barrier imposed by the dimensionality.

Nevertheless, due to periodic boundary conditions and to the system dimensionality increase, the correlations among the distances (triangular inequality, for example) are weakened so that in the high-dimensionality limit ($d \rightarrow \infty$), the distances between any two points can be considered as $N(N-1)/2$ i.d.d. random variables.
Thus, the RPP converges to the RLM. 

\section{Random Link Model}
\label{sec_rlm}

To work numerically with the RLM, one must generate directly the i.i.d. random distances. 
This solves the large computer RAM allocation problem due to high system dimensionality.
Because of the coordinate matrix inexistence, it is impossible to use the mask as in the finite dimensional systems (RPP). 
But, keeping track of the symmetry restriction ($D_{ij} = D_{ji}$) imposes serious numerical difficulties so that the distance matrix must be declared.
The symmetry restriction limits the RLM use to computational small systems.
For standard memory allocation at disposal, systems can have up to $N = 10^3$ points.

\subsection{Conventional Implementation}

A straightforward implementation of the i.i.d. random distances in RLM is schematized as follows.  
All the values in distance matrix main diagonal are null and only one seed $S_1$ is used to sequentially generate the distances of the main diagonal right-hand side (the distances on the diagonal left-hand side are obtained from  $D_{ij} = D_{ij}$).
The time dependence and memory consumption to run this computer algorithm are both proportional to $N^2$. 
(See Table.~\ref{table1})


We present below two methods, which replace the distance matrix by a mask, just as in the RPP. 
To re-obtain the distance $D_{ij}$ (which obeys the symmetry restriction), the deterministic feature of pseudo-random number generator is extensively explored. 

\subsection{One-Seed Method}

The first method is called \emph{single seed method} and reproduce any distance simply initializing the generator with the seed $S_1$ and calling the generator a given number of times. 
The only difficulty of this implementation is to introduce and control a counter to keep track of the exact number of calls to the random-number generator. 
The number of needed new random variables is $N-j$ for the $j$th row of the distance matrix. 
At each new row, this value must be added to the counter.

\begin{figure}
\begin{center}
\includegraphics[angle=-90, width = \columnwidth]{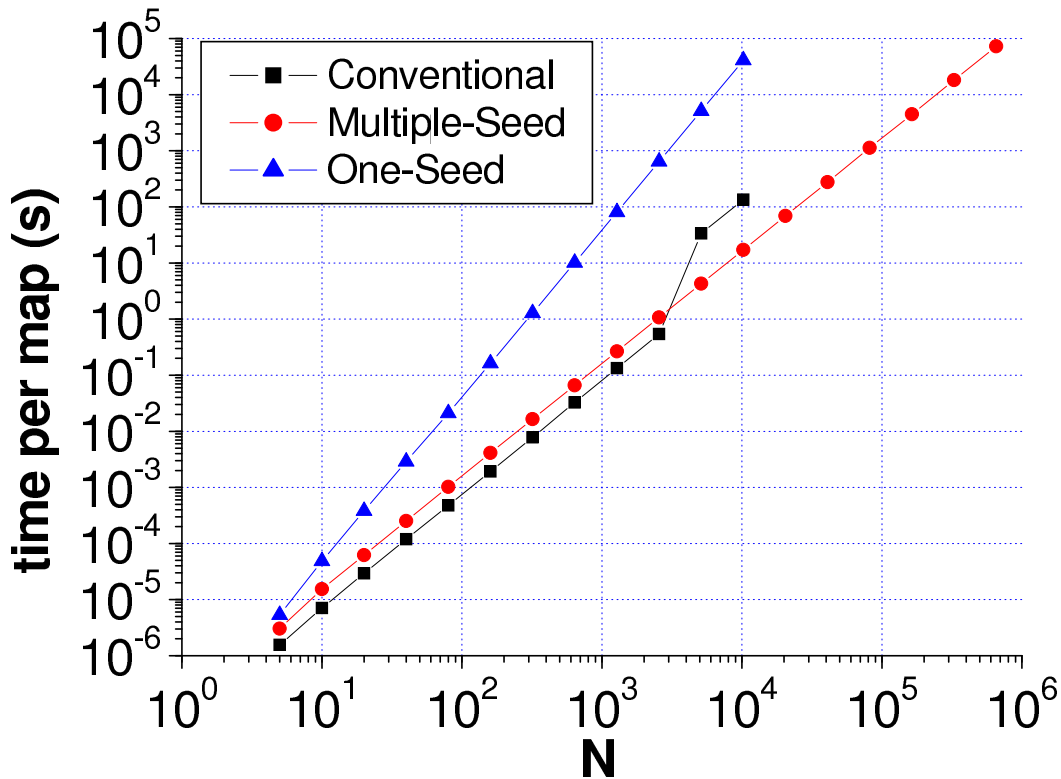}
\end{center}
\caption{Processing time as a function of system size. 
This dependence is well described by a power law with exponent $2$ for the conventional and multiple-seed methods and exponent $3$ for the one-seed method.  
While the conventional method can deal with systems of size around $10^3$, the multiple-seed method can deal with systems more than $100$ greater.
The discontinuity in the conventional method curve corresponds to the moment when swap to the disk started to be performed.}
\label{fig:time_cons}
\end{figure}

This method enables us to numerically construct much larger systems, but at the expense of a much longer computational time. 
To generate all distances in a $N$-point map, the memory consumption is of order $N$ due to the vector allocation (See Table~\ref{table1}) while required time is proportional to $N^3$ due to cumulative number of calls to the generator (See Figure~\ref{fig:time_cons}).

\begin{table}[htb]
\begin{center}
\begin{tabular}{cc|ccc}
\hline
        &Method& Conventional & One-Seed    & Multiple-Seed \\
\hline
Time    &         &   $N^2$  & $N^3$    & $N^2$         \\
Memory  &         &   $N^2$  & $N$      & $N$           \\
\hline
\end{tabular}
\end{center}
\caption{Memory and time consumptions for the conventional, one seed and multiple seed methods described. 
The multiple-seed method is the best one since it combines the low O$(N)$ memory consumption of the one-seed method and the low O$(N^2)$ time spent of the conventional implementation method.}
\label{table1}
\end{table}

\subsection{Multiple-Seed Method}

\begin{figure}[htb]
\begin{center}
\setlength{\unitlength}{1mm}

\begin{picture}(100,75)(0,30)

\multiput(0,100)(0,-10){6}{\line(1, 0){63}}
\multiput(0,100)(15,0){5}{\line(0, -1){53}}

\multiput(90,100)(0,-10){6}{\line(-1, 0){18}}
\multiput(90,100)(-15,0){2}{\line(0, -1){53}}

\multiput(0,100)(15,-10){4}{\makebox(15,-10)[c]{$0$}}

\put(15,100){\makebox(15,-10)[c]{$S_1$}}
\put(30,100){\makebox(15,-10)[c]{$S_2$}}
\put(45,100){\makebox(15,-10)[c]{$S_3$}}
\put(60,100){\makebox(15,-10)[c]{$\cdots$}}
\put(75,100){\makebox(15,-10)[c]{$S_{N-1}$}}
\put(90,100){\makebox(15,-10)[c]{$S_N$}}

\multiput(28,95)(15,0){5}{\vector(1, 0){4}}

\put(0,90){\makebox(15,-10)[c]{$S_1$}}
\put(0,80){\makebox(15,-10)[c]{$S_2$}}
\put(0,70){\makebox(15,-10)[c]{$S_3$}}
\put(0,60){\makebox(15,-10)[c]{$S_4$}}
\put(0,50){\makebox(15,-10)[c]{$\vdots$}}

\multiput(8,82)(0,-10){4}{\vector(0, -1){4}}

\put(30,90){\makebox(15,-10)[c]{$S_N$}}
\put(45,90){\makebox(15,-10)[c]{$S_{N+1}$}}
\put(60,90){\makebox(15,-10)[c]{$\cdots$}}
\put(75,90){\makebox(15,-10)[c]{$S_{2N-3}$}}
\put(90,90){\makebox(15,-10)[c]{$S_{2N-2}$}}

\multiput(43,85)(15,0){4}{\vector(1, 0){4}}

\put(15,80){\makebox(15,-10)[c]{$S_N$}}
\put(15,70){\makebox(15,-10)[c]{$S_{N+1}$}}
\put(15,60){\makebox(15,-10)[c]{$S_{N+2}$}}
\put(15,50){\makebox(15,-10)[c]{$\vdots$}}

\multiput(23,72)(0,-10){3}{\vector(0, -1){4}}

\put(45,80){\makebox(15,-10)[c]{$S_{2N-2}$}}
\put(60,80){\makebox(15,-10)[c]{$\cdots$}}
\put(75,80){\makebox(15,-10)[c]{$S_{3N-6}$}}
\put(90,80){\makebox(15,-10)[c]{$S_{3N-5}$}}

\multiput(58,75)(15,0){3}{\vector(1, 0){4}}

\put(30,70){\makebox(15,-10)[c]{$S_{2N-2}$}}
\put(30,60){\makebox(15,-10)[c]{$S_{2N-1}$}}
\put(30,50){\makebox(15,-10)[c]{$\vdots$}}

\multiput(38,62)(0,-10){2}{\vector(0, -1){4}}

\put(45,50){\makebox(15,-10)[c]{$\vdots$}}


\put(60,70){\makebox(15,-10)[c]{$\cdots$}}
\put(75,50){\makebox(15,-10)[c]{$\vdots$}}

\end{picture}
\caption{Multiple seed method scheme that stresses the seed evolution and the way they are stored.}
\label{fig:1}
\end{center}
\end{figure}
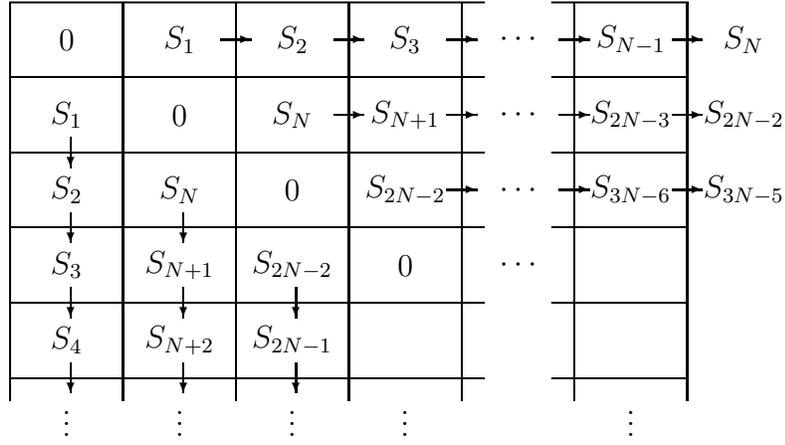

An improvement to the single seed method is obtained by the \emph{multiple seed method}. 
This method works noticing the following steps. 
Along the distance matrix first row, with the exception $D_{1,1}=0$, all other distances are new i.i.d. random variables and are generated simply making successive calls to the pseudo-random number generator.
Along the following rows, the distances on the main diagonal right-hand side are also new (i.i.d.) random variables and are generated in the same way as before.
Nevertheless, due to the model symmetry, the distances on the main diagonal left-hand side are not new random variables. 
Thus, the same seed $S$, which has been used to sequentially generate all the distances on the main diagonal right-hand side in the $k$-th row, must be used to sequentially generate all the distances below the main diagonal in the $k$-th column (See Figure~\ref{fig:1}).
For these requirements to be satisfied, the proposed method makes use of an integer vector of size $N$ (\emph{seed vector}) to store the seeds $S_1$,~$S_N$,~$S_{2N-2}$~$\ldots$, to be used to generate the distances in each one of the $N$ columns.   
Indeed, this vector maximum size is $N-1$, but using $N$ avoids undesired boundary tests. 
At every new row, these seeds are used to generate the distances on the left-hand side (and their new values -- modified by the generator -- are stored back in the seed vector) and the distances on the right-hand side are generated by simply making successive calls to the pseudo-random number generator (and only the first seed is added to the seed vector).

To implement these ideas, consider building the first row of the distance matrix.
Store the initial $S_1$ in the first entry of the seed vector (to use it further in the second row) and generate sequentially the $N-1$ random distances in the first row, after that temporarily save the last seed $S_N$. 
To generate the distance matrix second row, first use the stored $S_1$ seed from the seed vector to re-obtain the distance $D_{21}= D_{12}$, and store further two seeds $S_2$ and $S_N$ in the first two entries of the seed vector. 
Again, generate the following $N-2$ seeds and random distances in the second row saving the seed $S_{2N-2}$. 
The matrix distance third row is generated by the same procedure.
Use $S_2$ and $S_N$ to update the first two entries of the seed vector and add $S_{2N-2}$ to the third entry. 
This procedure is repeated successively up to the ($N-1$)th row.

Attention must be paid to a particular aspect.
In more sophisticated random number generators, internal static variables must be passed as the routine argument to obtain the desired reproducibility of the random-number generator.
Even in this case, the memory and time consumption orders of magnitude are not altered. 

\section{Conclusion}
\label{sec_conc}

Keeping track of the seeds along the construction of the all pair distances may drastically reduce computation time to order $N^{2}$ just like the conventional implementation, but with memory consumption of order $N$, just as the one-seed method. 
In this way, the algorithm presented here is the best compromise between time and memory consumptions.

\section{Acknowledgements}

We thank O. Kinouchi and R. Gonzalez Silva for fruitful discussions. 
A.S.M. acknowledges the support from the Brazilian agency CNPq (305527/2004-5).

\bibliographystyle{elsart-num}

\begin{thebibliography}{10}
\expandafter\ifx\csname url\endcsname\relax
  \def\url#1{\texttt{#1}}\fi
\expandafter\ifx\csname urlprefix\endcsname\relax\def\urlprefix{URL }\fi

\bibitem{mezard:1986}
M.~M\'ezard, G.~Parisi, Mean-field equations for the matching and travelling
  salesman problem, Europhys. Lett. 2~(12) (1986) 913--918.

\bibitem{harris:1960}
B.~Harris, Probability-distributions related to random mappings, Ann. Math.
  Stat. 31~(4) (1960) 1045--1062.

\bibitem{derrida:2:1997}
B.~Derrida, H.~Flyvbjerg, The random map model - a disordered model with
  deterministic dynamics, J. Phys. (Paris) 48~(6) (1987) 971--978.

\bibitem{kauffman:1969}
S.~A. Kauffman, Metabolic stability and epigenesis in randomly constructed
  genetic nets, J. Teor. Biol. 22 (1969) 437.

\bibitem{percus:1996}
A.~G. Percus, O.~C. Martin, Finite size and dimensional dependence in the
  {E}uclidean travelling salesman problem, Phys. Rev. Lett. 76~(8) (1996)
  1188--1191.

\bibitem{percus:1997}
N.~J. Cerf, J.~H.~B. de~Monvel, O.~Bohigas, O.~C. Martin, A.~G. Percus, The
  random link approximation for the {E}uclidean travelling salesman problem, J.
  Phys. I (France) 7 (1997) 117--136.

\bibitem{percus:1999}
A.~G. Percus, O.~C. Martin, The stochastic traveling salesman problem: finite
  size scaling and the cavity prediction, J. Stat. Phys. 94~(5-6) (1999)
  739--758.

\bibitem{aldous:2003}
D.~Aldous, A.~G. Percus, Scaling and universality in continous length
  combinatorial optimization, Proc. Nat. Acad. Sci. USA 100~(20) (2003)
  11211--11215.

\bibitem{aldous:2005}
D.~J. Aldous, Percolation-like scaling exponents for minimal paths and trees in
  stochastic mean field model, Proc. R. Soc. A 461 (2005) 825--838.

\bibitem{Boutet:1997}
J.~H.~B. de~Monvel, O.~C. Martin, Mean field and corrections for the
  {E}uclidean minimum matching problem, Phys. Rev. Lett. 79~(1) (1997)
  167--170.

\bibitem{Boutet:1998}
J.~Houdayer, J.~H.~B. de~Monvel, O.~C. Martin, Comparing mean field and
  {E}uclidean matching problems, Eur. Phys. J. B 6 (1998) 383--393.

\bibitem{kinouchi:1:2002}
O.~Kinouchi, A.~S. Martinez, G.~F. Lima, G.~M. Lourenço, S.~Risau-Gusman,
  Deterministic walks in random networks: an application to thesaurus graphs,
  Physica A 315~(3/4) (2002) 665--676.

\bibitem{tercariol_2005}
C.~A.~S. Ter\c{c}ariol, A.~S. Martinez, Analytical results for the statistical
  distribution related to memoryless deterministic tourist walk: Dimensionality
  effect and mean field models, Phys. Rev. E 72 (2005) 021103.

\bibitem{risaugusman:1:2003}
S.~Risau-Gusman, A.~S. Martinez, O.~Kinouchi, {E}scaping from cycles through a
  glass transition, Phys. Rev. E 68 (2003) Art. No. 016104 Part 2.

\bibitem{martinez:1:2004}
A.~S. Martinez, O.~Kinouchi, S.~Risau-Gusman, {E}xploratory behavior, trap
  models and glass transitions, Phys. Rev. E 69 (2004) Art. No. 017101 Part 2.

\bibitem{lima_prl2001}
G.~F. Lima, A.~S. Martinez, O.~Kinouchi, Deterministic walks in random media,
  Phys. Rev. Lett. 87~(1) (2001) 010603.

\bibitem{stanley_2001}
H.~E. Stanley, S.~V. Buldyrev, Statistical physics - the salesman and the
  tourist, Nature (London) 413~(6854) (2001) 373--374.

\bibitem{boyer_2004}
D.~Boyer, O.~Miramontes, G.~Ramos-Fernandez, J.~L. Mateos, G.~Cocho, Modeling
  the searching behavior of social monkeys, Physica A 342~(1-2) (2004)
  329--335.

\bibitem{boyer_2005}
D.~Boyer, H.~Larralde, Looking for the right thing at the right place: Phase
  transition in an agent model with heterogeneous spatial resources, Complexity
  10~(3) (2005) 52--55.

\end{thebibliography}

\end{document}